\documentstyle[aps,epsf,floats]{revtex}
\begin{document}
\draft
\twocolumn[\hsize\textwidth\columnwidth\hsize\csname 
           @twocolumnfalse\endcsname
\title{Quadrupole moments of rotating neutron stars}
\author{William G.~Laarakkers and Eric Poisson}
\address{Department of Physics, University of Guelph, Guelph,
         Ontario, N1G 2W1, Canada}
\date{Revised version, February 23, 1998}
\maketitle

\begin{abstract}
\widetext
Numerical models of rotating neutron stars are constructed for four 
equations of state using the computer code {\tt rns} written by 
Stergioulas. For five selected values of the star's gravitational 
mass (in the interval between 1.0 and 1.8 solar masses) and for 
each equation of state, the star's angular momentum is varied from 
$J=0$ to the Keplerian limit $J=J_{\rm max}$. For each neutron-star 
configuration we compute $Q$, the quadrupole moment of the mass 
distribution. We show that for given values of $M$ and $J$, $|Q|$ 
increases with the stiffness of the equation of state. For fixed
mass and equation of state, the dependence on $J$ is well reproduced 
with a simple quadratic fit, $Q \simeq - aJ^2/M c^2$, where $c$ is 
the speed of light, and $a$ is a parameter of order unity depending 
on the mass and the equation of state. 
\end{abstract}
\pacs{Subject headings: stars: neutron --- stars: rotation}
\vskip 2pc]

\narrowtext

\section{Introduction}

A rapidly rotating neutron star is not spherically symmetric, 
because the rotation creates a deformation in the stellar mass 
distribution. In turn, the star's oblateness creates a distortion 
in the gravitational field outside the star, which is measured by 
$Q_{ab}$, the quadrupole-moment tensor. Our purpose in this paper 
is to compute the quadrupole moments of rotating neutron stars, 
under the assumptions of rigid rotation, axial symmetry, and
reflection symmetry about the equatorial plane.

In Newtonian theory, the gravitational field outside any nonspherical 
body is given by (Jackson 1975)
\begin{equation}
\Phi(\vec{x}) = - G\, \frac{M}{r} - \frac{3G}{2}\, 
\frac{Q_{ab} x_a x_b}{r^5} + O(1/r^4),
\label{1}
\end{equation}
where $M$ is the body's mass and $r = |\vec{x}| = 
(\delta_{ab} x_a x_b)^{1/2}$ is the distance from the
center of mass; summation over repeated indices is understood. 
In terms of $\varepsilon$, the mass density inside 
the body, the quadrupole-moment tensor is given by
\begin{equation}
Q_{ab} = \int \varepsilon(\vec{x}')\, (x'_a x'_b - 
{\textstyle \frac{1}{3}} \delta_{ab} r'^2)\, d^3 x'.
\label{2}
\end{equation}
Thus, in Newtonian theory it is possible to define the 
quadrupole-moment tensor either in terms of the falloff behavior
of the gravitational potential, or in terms of an integral over 
the mass density.

If the body is axially symmetric about the $z$ direction, then 
Eqs.~(\ref{1}) and (\ref{2}) reduce to
\begin{equation}
\Phi(r,\theta) = - G\, \frac{M}{r} - 
G\, \frac{Q\, P_2(\cos\theta)}{r^3} + O(1/r^4)
\label{3}
\end{equation}
and
\begin{equation}
Q_{ab} = {\rm diag}(-{\textstyle \frac{1}{3}} Q,
-{\textstyle \frac{1}{3}} Q, {\textstyle \frac{2}{3}} Q),
\label{4}
\end{equation}
where $\cos\theta = z/r$, $P_2(x) = \frac{1}{2}(3x^2-1)$, and
\begin{equation}
Q = \int \varepsilon(r',\theta')\, r'^2\, P_2(\cos\theta')\, d^3x' .
\label{5}
\end{equation}
If the body is also reflection symmetric about the equatorial plane
($\theta=\pi/2$), then the $O(1/r^4)$ term vanishes identically and
Eq.~(\ref{3}) is valid up to terms of order $1/r^5$.

It is convenient to introduce the dimensionless quantity $q$, 
related to $Q$ and $M$ by
\begin{equation}
q = \frac{c^4}{G^2}\, \frac{Q}{M^3}.
\label{6}
\end{equation}
If we also introduce a dimensionless potential $\Phi^* \equiv \Phi/c^2$
and a characteristic length $M^* \equiv GM/c^2$, then Eq.~(\ref{3})
becomes $\Phi^*(r,\theta) = -M^*/r - q (M^*/r)^3 P_2(\cos\theta) + 
O(1/r^4)$. This equation resembles more closely the relativistic
analogue displayed in the following section. 

Because neutron stars are compact objects with strong internal
gravity, their gravitational fields must be described within the 
framework of general relativity. Equations (\ref{1})--(\ref{5})
are therefore not valid for neutron stars. In Sec.~2 of this 
paper we describe an operational way, due mostly to Fintan Ryan
(1995), of defining the quadrupole moment of an axially 
symmetric body in general relativity\footnote{In general relativity,
multipole moments are defined for both the mass density $\varepsilon$ 
and the current density $\vec{\j} = \varepsilon \vec{v}$, where 
$\vec{v}$ is the fluid velocity. Throughout this paper, the term 
``quadrupole moment'' will refer specifically to the quadrupole 
moment of the mass distribution.}. In Sec.~3 we describe a method 
for computing the quadrupole moment of a rotating neutron 
star, under the assumptions of rigid rotation, axial symmetry, and
reflection symmetry about the equatorial plane. Our computations 
rely on a numerical integration of the hydrostatic and Einstein field 
equations for selected equations of state, carried out with a numerical 
code written and made publicly available by Nikolaos Stergioulas (1995). 
In Sec.~4 we present and discuss our results. 

Computations of quadrupole moments of rotating neutron stars were
presented before in the literature: Datta (1988) has carried out 
such computations for a wide selection of equations of states, but
his results are restricted by an assumption of slow rotation; on
the other hand, Salgado {\it et al.}~(1994a and 1994b) have computed,
also for many equations of state, quadrupole moments for maximum-mass
configurations of rapidly-rotating neutron stars. While our own analysis 
does not incorporate a very large selection of equations of state, it 
explores more thoroughly the relationship between quadrupole moment 
and angular momentum, both for slow and rapid rotations. 

The motivation for this paper comes from the realization that the 
gravitational waves emitted by a binary system of rotating neutron 
stars will be affected by the stars' quadrupole moments. Although
this effect is small, it is comparable in magnitude to that due to 
the general relativistic spin-spin interaction (Kidder, Will, \& 
Wiseman, 1993; Apostolatos et al.~1994). This application of our 
results is discussed in a separate paper (Poisson 1998).

\section{Quadrupole moments of axially symmetric bodies in
         general relativity}

Equation (\ref{3}) states that the quadrupole moment $Q$ is the 
coefficient of the ``$P_2(\cos\theta)/r^3$'' part of the Newtonian 
potential. As we shall see, a similar statement holds in general 
relativity. However, because of the nonlinearities of the Einstein 
field equations, it is more difficult to express the quadrupole moment
in terms of an integral over the source. We shall come back to this
point in Sec.~3.

The metric of a stationary, axially symmetric body can be written
in the form (Bardeen \& Wagoner 1971)
\begin{eqnarray}
ds^2 &=& -e^{2\nu}\, dt^2 + r^2 \sin^2\theta B^2 e^{-2\nu} 
(d\phi - \omega\, dt)^2 
\nonumber \\ & & \mbox{}
+ e^{2\alpha} (dr^2 + r^2\, d\theta^2),
\label{7}
\end{eqnarray}
where the potentials $\nu$, $B$, $\omega$, and $\alpha$ are 
functions of $r$ and $\theta$. Butterworth and Ipser (1976) 
have calculated the asymptotic behavior of these potentials 
for large $r$. Apart from a slight change in notation, they find
\begin{eqnarray}
\nu &=& -M/r + {\textstyle \frac{1}{3}} b (M/r)^3  
\nonumber \\ & & \mbox{} 
- q (M/r)^3\, P_2(\cos\theta) 
+ O(1/r^4), 
\label{8} \\
B &=& 1 + b (M/r)^2 + O(1/r^4),
\label{9} \\
M\omega &=& 2\chi (M/r)^3 + O(1/r^4).
\label{10}
\end{eqnarray}
An expression for $\alpha$ will not be needed. Here, 
$M \equiv GM/c^2$ is the mass of the body, and $\chi$ is 
a dimensionless measure of its angular momentum:
\begin{equation}
\chi = \frac{J}{M^2} \equiv \frac{c}{G}\, \frac{J}{M^2},
\label{11}
\end{equation}
where $J$ is the body's angular momentum; as these equations 
indicate, we work with units such that $G=c=1$. The factors
of $G$ and $c$ will be re-inserted in Sec.~4.

Equations (\ref{8})--(\ref{10}) involve two additional (dimensionless) 
parameters, $b$ and $q$. The parameter $b$ was left undetermined 
by Butterworth and Ipser (1976). However, it is not a free parameter, 
and it is possible to calculate it by taking the spherically symmetric 
limit of the metric (\ref{7})--(\ref{10}) and comparing with the 
Schwarzschild solution in isotropic coordinates (Schutz 1985). Since 
these must be identical, we find 
\begin{equation}
b = - \frac{1}{4}.
\label{12}
\end{equation}
On the other hand, $q$ is a free parameter. Because $qM^3$ is the
coefficient of the ``$P_2(\cos\theta)/r^3$'' part of $\nu$ (which
may be identified as the Newtonian potential), this is the quantity
that should be interpreted as the quadrupole moment. The parameter 
$q$, then, would be identified with the dimensionless quantity 
introduced in Eq.~(\ref{6}).

The difficulty with this argument is that it is based upon the 
specific coordinate system of Eq.~(\ref{7}), and the question 
arises as to how it would be affected by a coordinate transformation. 
As we shall see, the statement that $qM^3$ is the quadrupole moment 
of the body is true irrespective of the coordinate system. The rest
of this section is devoted to a proof of this important statement.

There exists a well developed literature on coordinate-invariant
characterizations of multipole moments in general relativity. Such
work was pioneered by Geroch (1970) and pursued by Hansen (1974)
and Thorne (1980). Here we follow the simple prescription due to 
Ryan (1995), which is built on previous work by Fodor, Hoenselaers, 
and Perj\'es (1989). Ryan's prescription relies on the existence of 
a general relationship between the set of coordinate-invariant multipole 
moments defined by Hansen (1979), and a coordinate-invariant quantity 
$\Delta \tilde{E}$ defined as follows. 

Let a test particle move, in the absence of any external force, 
around a central body of mass $M$ and angular momentum $J$, in such 
a way that $\Omega$, its angular velocity as measured at infinity, is 
uniform. (Such orbits will be termed ``circular''.) The central body 
is assumed to be stationary, axially symmetric, and reflection symmetric
about the equatorial plane. We assume also that the motion of the test 
particle is confined to the equatorial plane, and that the particle moves 
in the same direction as the body's rotation. [If $\phi$ denotes the 
angle of rotation about the symmetry axis (which coincides with the body's 
rotation axis) and $t$ is the time measured at infinity, then $\Omega = 
d\phi/dt > 0$. Such coordinates are used in Eq.~(\ref{7}).] 

The orbital energy per unit test-particle mass, denoted $\tilde{E}$, 
is a coordinate-invariant quantity which depends on the orbital 
motion of the particle. Because each circular orbit can be labeled 
with the value of its angular velocity $\Omega$ (also a 
coordinate-invariant quantity), we have that $\tilde{E}$ can be
expressed as a function of $\Omega$. The quantity considered by 
Ryan (1995) is 
\begin{equation}
\Delta \tilde{E} \equiv - \Omega \frac{d\tilde{E}}{d\Omega}.
\label{13}
\end{equation}
This evidently is a coordinate-invariant quantity,
and Ryan derives the following expression for it:
\begin{eqnarray}
\Delta \tilde{E} &=& \frac{1}{3}\, v^2 - \frac{1}{2}\, v^4 + 
\frac{20}{9}\, \frac{J}{M^2}\, v^5 
\nonumber \\ & & \mbox{}
+ \biggl( \frac{Q}{M^3} - 
\frac{27}{8} \biggr)\, v^6 + O(v^7),
\label{14}
\end{eqnarray}
where $Q$ (denoted $M_2$ by Ryan) is the coordinate-invariant 
quadrupole moment, and
\begin{equation}
v \equiv (M\Omega)^{1/3}
\label{15}
\end{equation}
is the orbital velocity. Ryan's formula is valid for any spacetime
satisfying the assumptions listed above. The quadrupole moment of
a selected spacetime is therefore determined by computing 
$\Delta \tilde{E}$ explicitly for this spacetime, and comparing
with Eq.~(\ref{14}). Because all involved quantities are known to 
be invariant under a coordinate transformation, this calculation can 
be carried out in any coordinate system.

We now go through the steps of calculating $\Delta \tilde{E}$ for
the specific spacetime of Eqs.~(\ref{7})--(\ref{10}). 

Starting from the $r$ component of the geodesic equation, using 
the fact that $r$ and $\theta$ are both constant on circular,
equatorial orbits, we easily derive the following expression for 
the angular velocity:
\begin{equation}
\Omega = \frac{d\phi}{dt} 
= \frac{ -g_{t\phi,r} + \sqrt{ (g_{t\phi,r})^2 - 
g_{tt,r} g_{\phi\phi,r} } }{ g_{\phi\phi,r} },
\label{16}
\end{equation}
where a comma indicates partial differentiation, and all metric 
functions are evaluated at $\theta=\pi/2$. Reading off the metric 
components from Eq.~(\ref{7}), substituting the asymptotic relations 
(\ref{8})--(\ref{10}), taking a cubic root, and finally, expanding in 
powers of $x \equiv (M/r)^{1/2}$, yields
\begin{equation}
v = x - \frac{1}{2}\, x^3 - \frac{1}{3}\, \chi\, x^4 + 
\frac{1-q}{4}\, x^5 + O(x^6).
\label{17}
\end{equation}
This series can be inverted to give
\begin{equation}
x = v + \frac{1}{2}\, v^3 + \frac{1}{3}\, \chi\, v^4 +
\frac{2+q}{4}\, v^5 + O(v^6).
\label{18}
\end{equation}

An expression for $\tilde{E}$ follows from the fact that in the 
coordinates of Eq.~(\ref{7}), $\tilde{E} = -g_{t\alpha} u^\alpha$, 
where $u^\alpha = (u^t,0,0,u^\phi) = u^t(1,0,0,\Omega)$ is the 
particle's four-velocity. Using also the normalization condition
$g_{\alpha\beta} u^\alpha u^\beta = -1$, we arrive at
\begin{equation}
\tilde{E} = \frac{ -g_{tt} - g_{t\phi} \Omega }{\sqrt{
-g_{tt} - 2 g_{t\phi} \Omega - g_{\phi\phi} \Omega^2 }}.
\label{19}
\end{equation}
Going through the same steps as before, we obtain
\begin{equation}
\tilde{E} = 1 - \frac{1}{2}\, x^2 + \frac{7}{8}\, x^4 - 
\chi\, x^5 + \frac{9-4q}{16}\, x^6 + O(x^7),
\label{20}
\end{equation}
or, after substituting Eq.~(\ref{18}),
\begin{equation}
\tilde{E} = 1 - \frac{1}{2}\, v^2 + \frac{3}{8}\, v^4 -
\frac{4}{3}\, \chi\, v^5 + \frac{27-8q}{16}\, v^6 + O(v^7).
\label{21}
\end{equation}

Finally, Eqs.~(\ref{13}) and (\ref{15}) give $\Delta \tilde{E} 
= -(v/3) d\tilde{E}/dv$, and substituting Eq.~(\ref{21}), we
arrive at
\begin{equation}
\Delta \tilde{E} = \frac{1}{3}\, v^2 - \frac{1}{2}\, v^4 +
\frac{20}{9}\, \chi\, v^5 + \biggl( q - \frac{27}{8} \biggr)\, v^6
+ O(v^7).
\label{22}
\end{equation}
Comparing with the general relation (\ref{14}) confirms that
$J = \chi M^2$, as was first stated in Eq.~(\ref{11}), and
establishes that the quadrupole moment is given by $Q = qM^3$, 
in accordance with our original guess. 

We conclude that the dimensionless parameter $q$ appearing in 
Eq.~(\ref{8}) is related to the coordinate invariant multipole 
moment $Q$ by the relation 
\begin{equation}
Q = qM^3,
\label{23}
\end{equation}
which is formally identical to Eq.~(\ref{6}). We see that the 
quadrupole moment is indeed determined by isolating the 
``$P_2(\cos\theta)/r^3$'' part of $\nu$, the relativistic 
analogue of the Newtonian potential. 

\section{Computation of the quadrupole moment}

The computation of the quadrupole moment of a neutron star 
requires a computer code capable of solving the hydrostatic and 
Einstein field equations for uniformly rotating mass distributions,
under the assumptions of stationarity, axial symmetry about the
rotation axis, and reflection symmetry about the equatorial plane. 
Along with the density profile $\varepsilon(r,\theta)$ and other
matter variables, the code must compute the metric functions $\nu$, 
$B$, $\omega$, and $\alpha$. Such a code was recently written and 
made publicly available by Stergioulas (1995). [See Stergioulas \&
Friedman (1995) for a comparison with other codes.] The code, named 
{\tt rns}, uses tabulated equations of state for neutron-star matter, 
and is based upon the numerical methods of Komatsu, Eriguchi, and 
Hachisu (1989). As a technical point, we may remark that {\tt rns} 
uses the metric variables $\rho = 2\nu - \ln B$ and $\gamma = \ln B$ 
instead of $\nu$ and $B$. 

Once $\nu(r,\theta)$ has been obtained for a given equation of state 
and for selected values of stellar parameters, such as gravitational 
mass $M$ and angular momentum $J$, the computation of $q$ is in principle 
straightforward. Indeed, Eq.~(\ref{8}) immediately implies
\begin{equation}
q = - \frac{5}{2}\, \lim_{r \to \infty}\, (r/M)^3 \int_{-1}^{1}
\nu(r,\theta)\, P_2(\cos\theta)\, d\cos\theta ,
\label{24}
\end{equation}
and the integral to the right-hand side can easily be evaluated 
numerically. The difficulty with implementing this method is that 
when $r$ is much larger than $M$, the ``$P_2(\cos\theta)/r^3$'' 
part of $\nu$ is extremely small compared with its spherically 
symmetric part. The method therefore lacks numerical accuracy. 
Nevertheless, we have found Eq.~(\ref{24}) useful as a check on 
the results of another method of computing $q$, which we now 
describe.

This method was first devised by Salgado {\it et al.}~(1994a), and 
then independently by Ryan (1997), who shows that the quadrupole 
moment can be straightforwardly computed as an integral over 
$S_\rho(r',\theta')$, a complicated function involving matter
variables and metric components. This function is defined in Eq.~(10) 
of Komatsu, Eriguchi, and Hachisu (1989). Ryan's expression is
\begin{equation}
Q = \frac{1}{8\pi}\, \int S_\rho(r',\theta')\, r'^2\, 
P_2(\cos\theta')\, d^3 x'.
\label{25}
\end{equation}
This is the general relativistic analogue of Eq.~(\ref{5}). 
Although Eq.~(\ref{25}) looks quite simple, many complications 
associated with the nonlinearities of general relativity are 
hidden in the function $S_\rho$ which, unlike the mass density
$\varepsilon$, has support both inside and outside the star.
It is only in the Newtonian limit that $S_\rho = 8\pi \varepsilon$.
Nevertheless, this method is easy to implement because {\tt rns} 
computes $S_\rho$ and performs similar integrations in order 
to solve the Einstein field equations. It was therefore 
straightforward to make the necessary modifications to the 
code, and compute $q = Q/M^3$ for selected configurations of 
rotating neutron stars.

\section{Results and discussion}

With {\tt rns} we have constructed numerical models of rotating 
neutron stars for four different tabulated equations of state,
those labeled G, FPS, C, and L. [See Cook, Shapiro, and Teukolsky 
(1994) for a detailed description of these equations of state.] 
For five selected values of the gravitational mass (in the interval
between $1.0\ M_\odot$ and $1.8\ M_\odot$) and for each equation 
of state, we have constructed models of varying angular momentum 
$J$, from $J=0$ to the Keplerian limit $J=J_{\rm max}$, at which the 
star's angular velocity exceeds that of a test particle in circular, 
equatorial motion just outside the stellar surface. [Of the five 
selected values for the mass, $M=1.4\ M_\odot$ is probably the most 
relevant, because it is the most commonly observed for neutron stars 
in binary systems (Finn 1994).] For each of these neutron-star models 
we have computed the quadrupole moment $Q$ using the two methods 
described in the previous section.

\begin{table}
\caption{Properties of equations of state. The entries for 
$M_{\rm max}$ are taken from Cook, Shapiro, and Teukolsky (1994).}
\begin{tabular}{ccc}
EOS & $M_{\rm max}/M_\odot$ for $J=0$ & 
$\chi_{\rm max}$ for $M=1.4\, M_\odot$ \\
\hline 
G   & 1.36 & 0.616 \\
FPS & 1.80 & 0.669 \\
C   & 1.86 & 0.654 \\
L   & 2.70 & 0.731 
\end{tabular}
\end{table}

Table 1 displays some properties of the selected equations of
state. The first column lists the equations of state, in order
of increasing stiffness. The second column lists $M_{\rm max}$ 
(in solar masses) for $J=0$, the maximum value of the gravitational 
mass for a nonrotating configuration. The third column lists 
$\chi_{\rm max} = c J_{\rm max}/GM^2$ for $M=1.4\, M_\odot$. 
The first row of Table 1 indicates that EOS G is extremely soft, 
with a maximum mass (in the absence of rotation) {\it lower} than 
$1.4\, M_\odot$. The models constructed for this equation of state 
therefore belong to the supramassive class (Cook, Shapiro, \&
Teukolsky 1994), and are probably not realistic. Furthermore,
even rotation cannot support such a star if its mass is larger
than 1.56 solar masses (Cook, Shapiro, \& Teukolsky 1994).

Our results are presented in Tables 2--6. For each of the
equations of state, the first column lists $\chi = cJ/GM^2$, 
the dimensionless angular-momentum parameter, and the second 
column lists the corresponding value of $q = c^4 Q / G^2 M^3$, 
the dimensionless quadrupole-moment parameter. The last entry
corresponds to a configuration of maximum rotation. We recall
that neutron stars of 1.6 and 1.8 solar masses cannot be 
constructed with EOS G. As the number of digits quoted in the 
tables indicates, we believe our computations to be accurate at 
least to one part in $10^3$.

\begin{table}
\caption{Quadrupole moment: $M=1.0\ M_\odot$}
\begin{tabular}{cccccccc}
\multicolumn{2}{c}{EOS G} & \multicolumn{2}{c}{EOS FPS} &
\multicolumn{2}{c}{EOS C} & \multicolumn{2}{c}{EOS L} \\
\hline 
$\chi$ & $q$ & $\chi$ & $q$ & $\chi$ & $q$ & $\chi$ & $q$ \\
\hline 
 0.173 & -0.158 & 0.182 & -0.289 & 0.166 & -0.279 & 0.172 & -0.396 \\
 0.232 & -0.275 & 0.245 & -0.514 & 0.257 & -0.637 & 0.269 & -0.960 \\
 0.279 & -0.391 & 0.293 & -0.709 & 0.325 & -1.028 & 0.340 & -1.513 \\
 0.320 & -0.515 & 0.336 & -0.921 & 0.382 & -1.404 & 0.400 & -2.053 \\
 0.358 & -0.642 & 0.373 & -1.125 & 0.432 & -1.786 & 0.452 & -2.578 \\
 0.393 & -0.774 & 0.408 & -1.329 & 0.477 & -2.164 & 0.500 & -3.097 \\
 0.425 & -0.909 & 0.440 & -1.540 & 0.519 & -2.541 & 0.543 & -3.608 \\
 0.460 & -1.067 & 0.469 & -1.744 & 0.559 & -2.928 & 0.584 & -4.118 \\
 0.495 & -1.239 & 0.497 & -1.955 & 0.596 & -3.319 & 0.622 & -4.625 \\
 0.531 & -1.434 & 0.525 & -2.167 & 0.631 & -3.713 & 0.658 & -5.126 \\
 0.566 & -1.635 & 0.551 & -2.375 & 0.632 & -3.720 & 0.693 & -5.629 \\
 0.599 & -1.844 & 0.644 & -3.217 &       &        &       &        
\end{tabular}
\end{table}

\begin{figure}
\special{hscale=35 vscale=35 hoffset=-20.0 voffset=20.0
         angle=-90.0 psfile=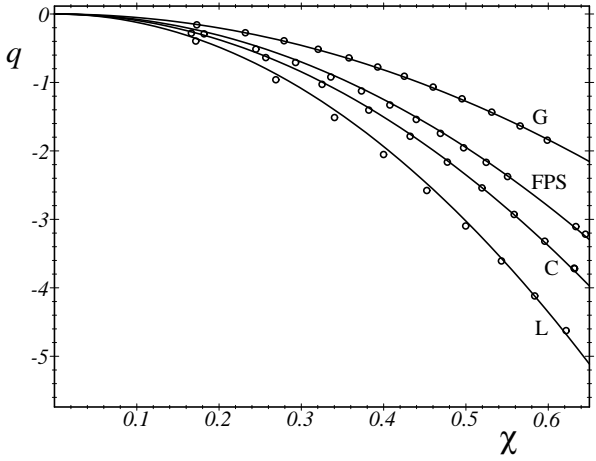}
\vspace*{2.6in}
\caption{Plots of $q$ as a function of $\chi$ for four 
equations of state, for neutron stars of mass $M=1.0\ M_\odot$.
The data of Table 2 is shown with circles, and the solid curves 
represent the fits of Eq.~(\ref{27}).}
\end{figure}

\begin{table}
\caption{Quadrupole moment: $M=1.2\ M_\odot$}
\begin{tabular}{cccccccc}
\multicolumn{2}{c}{EOS G} & \multicolumn{2}{c}{EOS FPS} &
\multicolumn{2}{c}{EOS C} & \multicolumn{2}{c}{EOS L} \\
\hline 
$\chi$ & $q$ & $\chi$ & $q$ & $\chi$ & $q$ & $\chi$ & $q$ \\
\hline 
 0.203 & -0.124 & 0.191 & -0.209 & 0.175 & -0.213 & 0.146 & -0.246 \\
 0.283 & -0.241 & 0.276 & -0.448 & 0.222 & -0.351 & 0.258 & -0.683 \\
 0.345 & -0.363 & 0.341 & -0.676 & 0.259 & -0.475 & 0.334 & -1.120 \\
 0.398 & -0.491 & 0.398 & -0.920 & 0.292 & -0.596 & 0.396 & -1.546 \\
 0.445 & -0.625 & 0.448 & -1.156 & 0.322 & -0.719 & 0.450 & -1.975 \\
 0.488 & -0.765 & 0.494 & -1.394 & 0.349 & -0.845 & 0.499 & -2.395 \\
 0.528 & -0.910 & 0.536 & -1.639 & 0.376 & -0.969 & 0.544 & -2.809 \\
 0.563 & -1.054 & 0.577 & -1.892 & 0.399 & -1.088 & 0.586 & -3.218 \\
 0.596 & -1.206 & 0.615 & -2.148 & 0.422 & -1.216 & 0.625 & -3.623 \\
       &        & 0.651 & -2.412 & 0.444 & -1.343 & 0.663 & -4.032 \\
       &        &       &        & 0.515 & -1.805 & 0.698 & -4.439 \\
       &        &       &        & 0.579 & -2.275 &       &        \\
       &        &       &        & 0.638 & -2.768 &       &         
\end{tabular}
\end{table}

\begin{figure}
\special{hscale=35 vscale=35 hoffset=-20.0 voffset=20.0
         angle=-90.0 psfile=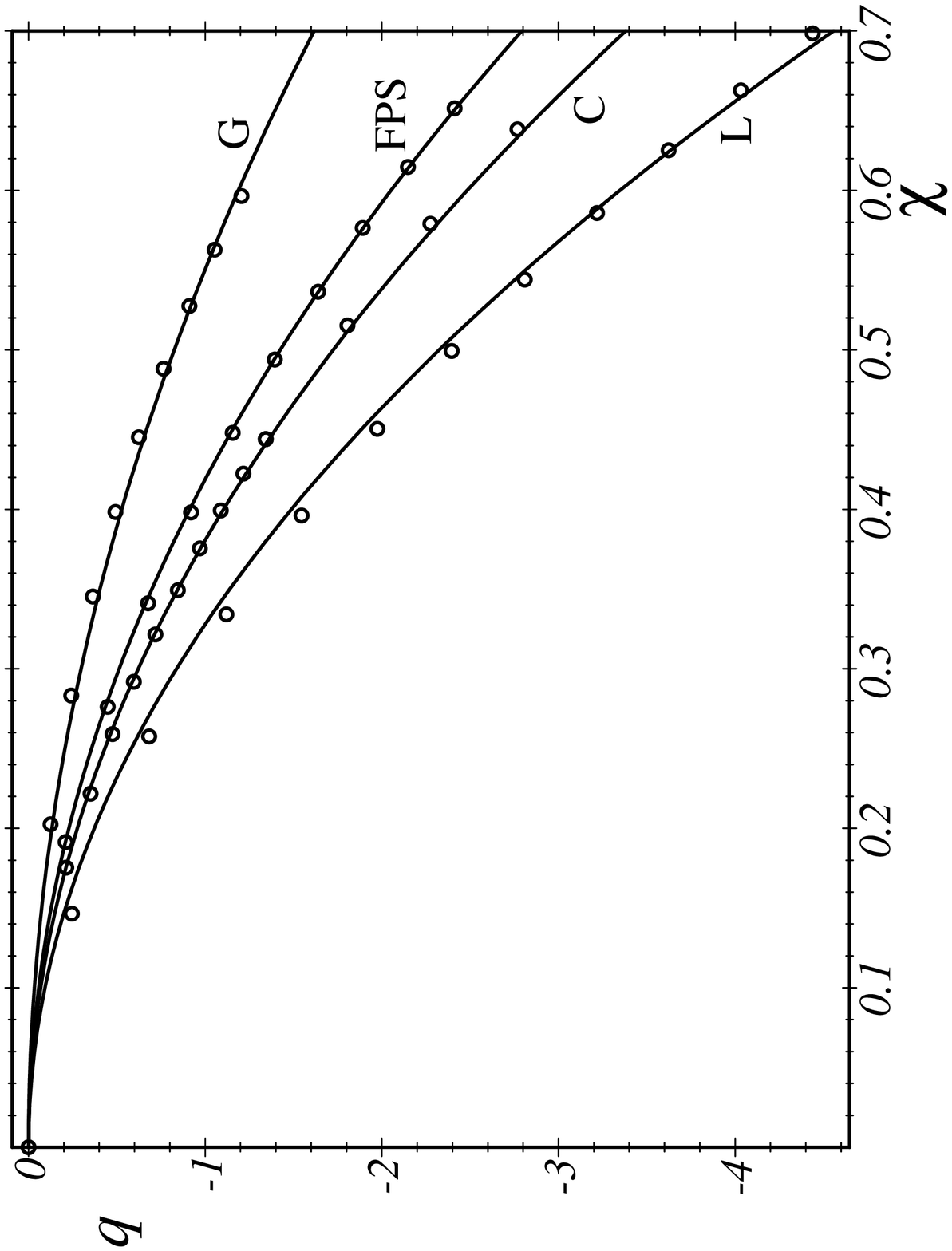}
\vspace*{2.6in}
\caption{Same as Fig.~1 but with $M=1.2\ M_\odot$. The
circles represent the data of Table 3.}
\end{figure}

\begin{table}
\caption{Quadrupole moment: $M=1.4\ M_\odot$}
\begin{tabular}{cccccccc}
\multicolumn{2}{c}{EOS G} & \multicolumn{2}{c}{EOS FPS} &
\multicolumn{2}{c}{EOS C} & \multicolumn{2}{c}{EOS L} \\
\hline 
$\chi$ & $q$ & $\chi$ & $q$ & $\chi$ & $q$ & $\chi$ & $q$ \\
\hline 
 0.337 & -0.173 & 0.119 & -0.074 & 0.200 & -0.199 & 0.124 & -0.134 \\
 0.354 & -0.196 & 0.224 & -0.216 & 0.269 & -0.356 & 0.234 & -0.454 \\
 0.372 & -0.223 & 0.292 & -0.355 & 0.325 & -0.523 & 0.308 & -0.764 \\
 0.393 & -0.256 & 0.347 & -0.501 & 0.373 & -0.690 & 0.368 & -1.070 \\
 0.416 & -0.295 & 0.396 & -0.655 & 0.417 & -0.861 & 0.420 & -1.373 \\
 0.440 & -0.341 & 0.440 & -0.808 & 0.457 & -1.036 & 0.467 & -1.672 \\
 0.466 & -0.394 & 0.481 & -0.970 & 0.494 & -1.211 & 0.510 & -1.975 \\
 0.493 & -0.456 & 0.519 & -1.134 & 0.529 & -1.392 & 0.551 & -2.279 \\
 0.521 & -0.526 & 0.556 & -1.302 & 0.562 & -1.578 & 0.589 & -2.587 \\
 0.549 & -0.605 & 0.591 & -1.476 & 0.593 & -1.768 & 0.626 & -2.894 \\
 0.578 & -0.696 & 0.625 & -1.659 & 0.623 & -1.962 & 0.661 & -3.199 \\
 0.606 & -0.795 & 0.658 & -1.847 & 0.653 & -2.166 & 0.694 & -3.507 
\end{tabular}
\end{table}

\begin{figure}
\special{hscale=35 vscale=35 hoffset=-20.0 voffset=20.0
         angle=-90.0 psfile=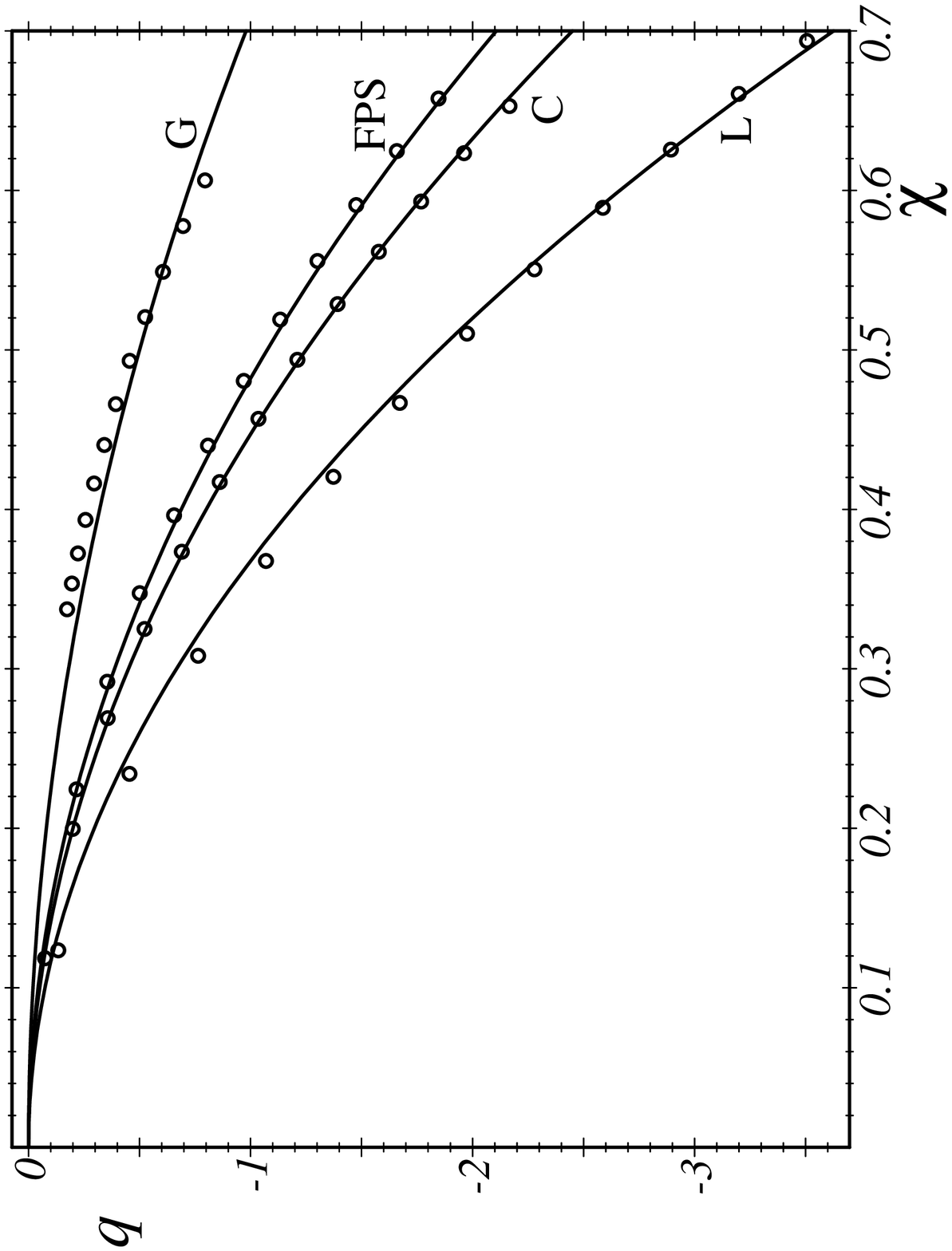}
\vspace*{2.6in}
\caption{Same as Fig.~1 but with $M=1.4\ M_\odot$. The
circles represent the data of Table 4.}
\end{figure}

\begin{table}
\caption{Quadrupole moment: $M=1.6\ M_\odot$}
\begin{tabular}{cccccc}
\multicolumn{2}{c}{EOS FPS} &
\multicolumn{2}{c}{EOS C} & \multicolumn{2}{c}{EOS L} \\
\hline 
$\chi$ & $q$ & $\chi$ & $q$ & $\chi$ & $q$ \\
\hline 
 0.111 & -0.037 & 0.130 & -0.066 & 0.131 & -0.120 \\
 0.225 & -0.158 & 0.223 & -0.184 & 0.246 & -0.388 \\
 0.299 & -0.267 & 0.286 & -0.289 & 0.324 & -0.672 \\
 0.361 & -0.390 & 0.340 & -0.409 & 0.387 & -0.944 \\
 0.415 & -0.517 & 0.386 & -0.528 & 0.442 & -1.216 \\
 0.464 & -0.652 & 0.429 & -0.658 & 0.491 & -1.488 \\
 0.509 & -0.797 & 0.468 & -0.793 & 0.537 & -1.760 \\
 0.552 & -0.953 & 0.506 & -0.937 & 0.580 & -2.033 \\
 0.594 & -1.120 & 0.543 & -1.089 & 0.620 & -2.310 \\
 0.635 & -1.297 & 0.578 & -1.248 & 0.658 & -2.589 \\
       &        & 0.612 & -1.417 & 0.695 & -2.872 \\
       &        & 0.645 & -1.596 & 0.731 & -3.162  
\end{tabular}
\end{table}

\begin{figure}
\special{hscale=35 vscale=35 hoffset=-20.0 voffset=20.0
         angle=-90.0 psfile=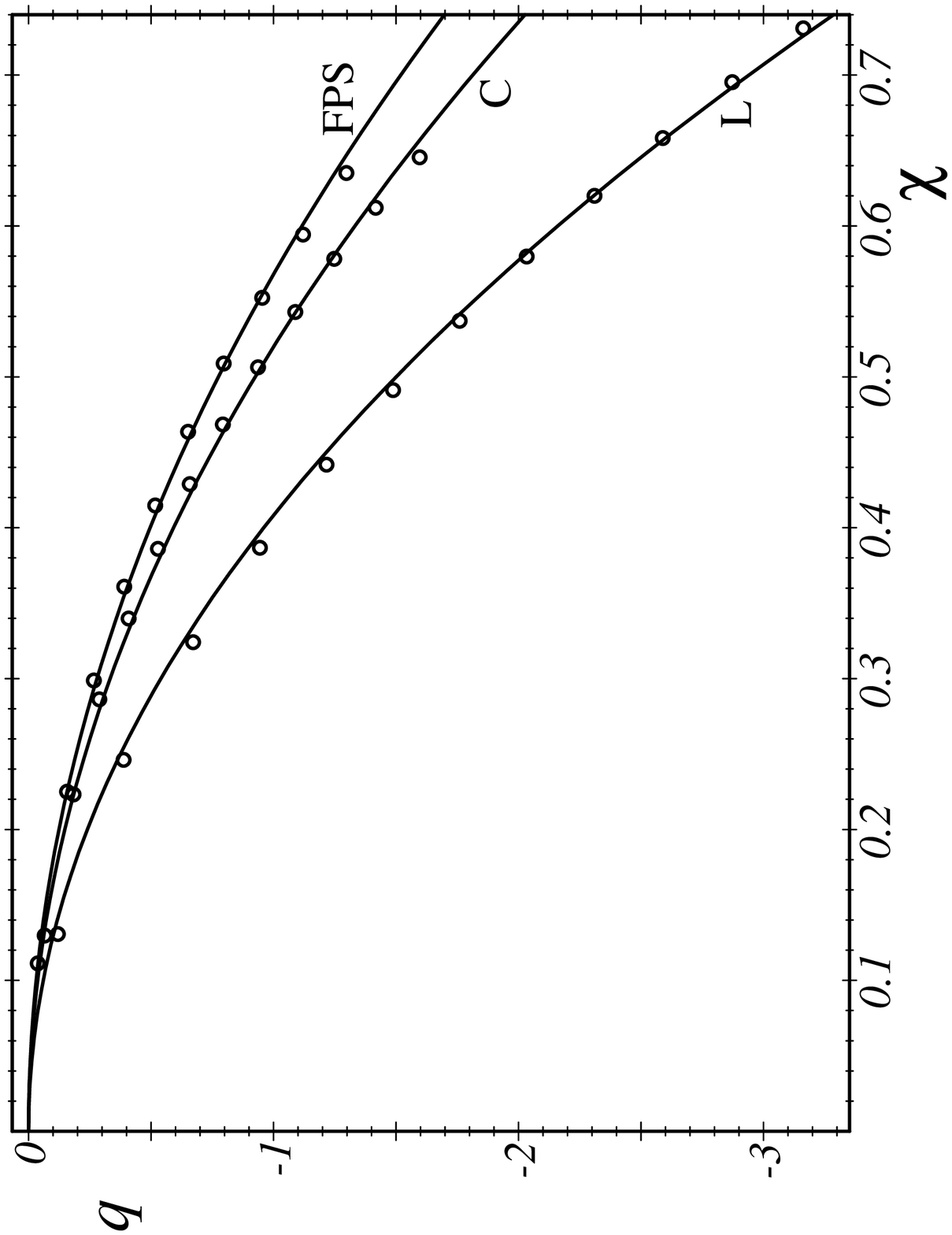}
\vspace*{2.6in}
\caption{Same as Fig.~1 but with $M=1.6\ M_\odot$. The
circles represent the data of Table 5.}
\end{figure}

\begin{table}
\caption{Quadrupole moment: $M=1.8\ M_\odot$}
\begin{tabular}{cccccc}
\multicolumn{2}{c}{EOS FPS} &
\multicolumn{2}{c}{EOS C} & \multicolumn{2}{c}{EOS L} \\
\hline 
$\chi$ & $q$ & $\chi$ & $q$ & $\chi$ & $q$ \\
\hline 
 0.113 & -0.024 & 0.112 & -0.030 & 0.136 & -0.109 \\
 0.169 & -0.052 & 0.191 & -0.081 & 0.250 & -0.326 \\
 0.222 & -0.091 & 0.253 & -0.152 & 0.326 & -0.549 \\
 0.274 & -0.143 & 0.304 & -0.222 & 0.388 & -0.768 \\
 0.323 & -0.203 & 0.353 & -0.306 & 0.443 & -0.990 \\
 0.372 & -0.278 & 0.399 & -0.396 & 0.491 & -1.214 \\
 0.420 & -0.365 & 0.442 & -0.496 & 0.537 & -1.438 \\
 0.466 & -0.464 & 0.483 & -0.607 & 0.579 & -1.664 \\
 0.513 & -0.579 & 0.524 & -0.732 & 0.619 & -1.893 \\
 0.559 & -0.711 & 0.564 & -0.871 & 0.657 & -2.128 \\
 0.605 & -0.863 & 0.604 & -1.026 & 0.694 & -2.367 \\
 0.654 & -1.046 & 0.645 & -1.202 & 0.729 & -2.612  
\end{tabular}
\end{table}

\begin{figure}
\special{hscale=35 vscale=35 hoffset=-20.0 voffset=20.0
         angle=-90.0 psfile=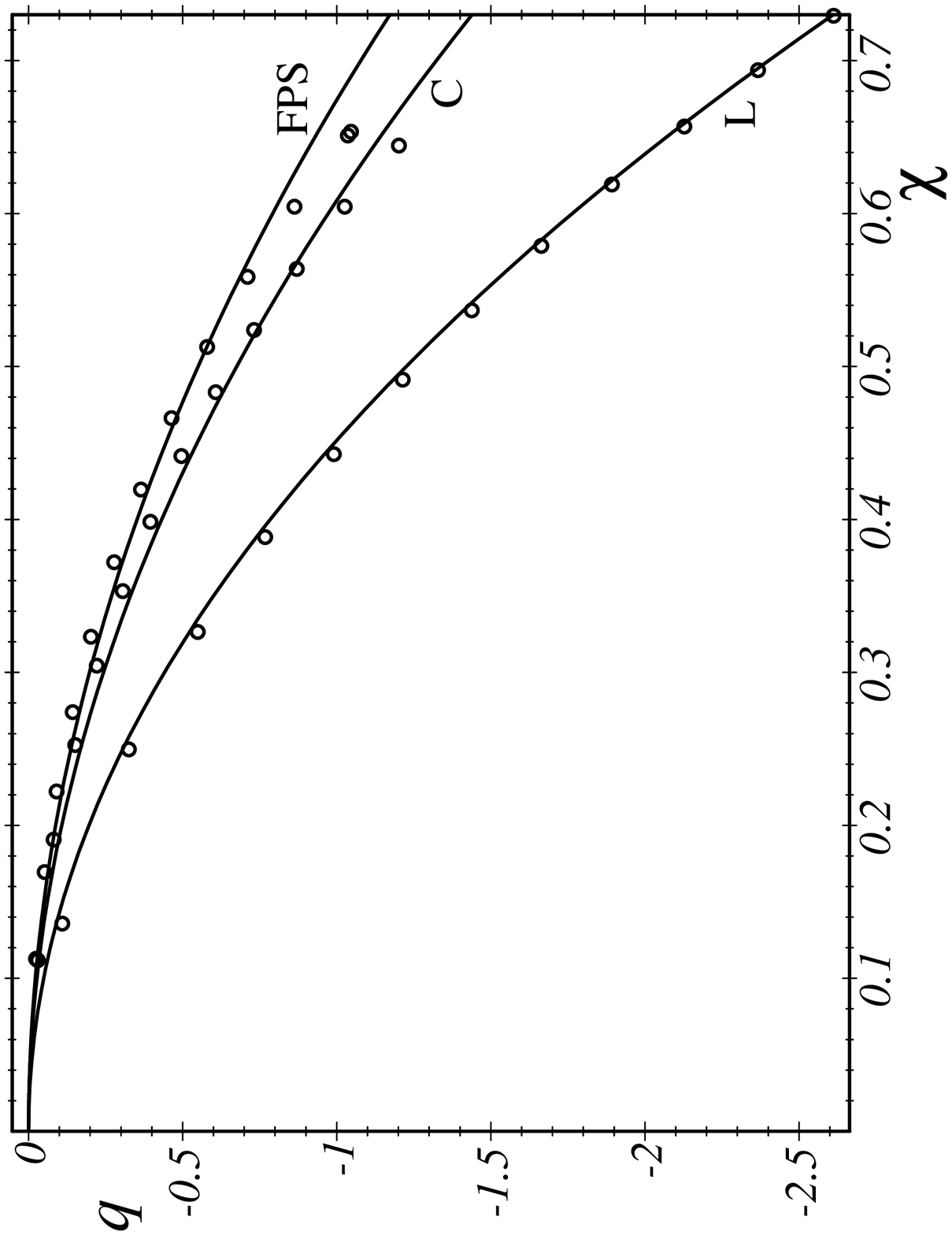}
\vspace*{2.6in}
\caption{Same as Fig.~1 but with $M=1.8\ M_\odot$. The
circles represent the data of Table 6.}
\end{figure}

The fact that the quadrupole moments are all negative reflects the 
oblateness of the mass distribution; the rotation produces a bulge 
at the equator. For given values of $M$ and $\chi$, we find that 
$|q|$ increases with the stiffness of the equation of state. Thus, 
$|q(\mbox{G})| < |q(\mbox{FPS})| < |q(\mbox{C})| < |q(\mbox{L})|$. 
This was to be expected, because a stiffer equation of state produces 
a larger star, and the quadrupole moment scales with the square of the 
star's radius. For fixed mass and equation of state, the dependence on 
$\chi$ is well reproduced with a simple quadratic fit,
\begin{equation}
q \simeq -a(M,\mbox{EOS})\, \chi^2
\label{27}
\end{equation}
or
\begin{equation}
Q \simeq -\frac{a(M,\mbox{EOS})}{c^2}\, \frac{J^2}{M},
\label{28}
\end{equation}
with the parameter $a$ depending on the mass and the equation of state. 
The best-fit values are presented in Table 7, which shows that for a
fixed equation of state, $a$ decreases with increasing mass. (The 
relation is not well reproduced by simple fitting formulas, such as 
linear and power-law relationships.) The accuracy of these fits is 
demonstrated in Figs.~1--5, which display $q$ as a function of $\chi$ 
for each of the selected equations of state. The figures show the data 
points of Table 2--6 as well as the fits. We see that Eq.~(\ref{27}) 
reproduces the data quite well.

\begin{table}
\caption{Fit parameter}
\begin{tabular}{ccccc}
$M/M_\odot$ & \multicolumn{4}{c}{$a$} \\
\hline
  & EOS G & EOS FPS & EOS C & EOS L \\  
\hline 
1.0 & 5.1 & 7.8 & 9.4 & 12.1 \\
1.2 & 3.3 & 5.7 & 6.9 &  9.3 \\
1.4 & 2.0 & 4.3 & 5.0 &  7.4 \\
1.6 &     & 3.1 & 3.7 &  6.0 \\
1.8 &     & 2.2 & 2.7 &  4.9
\end{tabular}
\end{table}

The high degree of accuracy of the formula $q \simeq -a \chi^2$ is 
intriguing, because it holds not only for slow rotations, where such a 
quadratic relation is to be expected, but also for fast rotations, where 
such a relation is expected to fail. Here, our expectation is based on the 
theory of uniformly-rotating, constant-density fluid configurations in 
Newtonian gravity (Chandrasekhar 1969). For such configurations, called 
Maclaurin spheriods, the relation between $q$ and $\chi$ is known exactly. 
For slow rotation, the relation is quadratic: $q \simeq -(25/8)(c^2 s_p/G M)\, 
\chi^2$, where $s_p$ is the polar radius. For fast rotations, however, 
the relation deviates strongly from a quadratic. For typical neutron-star
parameters ($M=1.4\ M_\odot$ and $s_p = 15\ \mbox{km}$), the deviation 
becomes significant when $\chi \sim 0.4$, which is {\it smaller} than 
the maximum value ($\chi_{\rm max} > 0.6$). Therefore, the reliability 
of the relation (\ref{27}) in the complete interval $0 < \chi < \chi_{\rm max}$ 
is quite remarkable. In this regard, we may note that the slow-rotation 
approximation was shown by Weber and Glendenning (1992) to give rather 
accurate results in numerical models of rotating neutron stars. We also 
note that the relation $q=-\chi^2$ is known to hold {\it exactly} for 
rotating black holes (Thorne 1980). 

\acknowledgments

This work was supported by the Natural Sciences and Engineering Research 
Council of Canada. It is a pleasure to thank Fintan Ryan for a very helpful 
exchange of e-mails regarding the computation of multipole moments in general 
relativity, and Curt Cutler for explaining why stiffer equations of state
produce larger quadrupole moments. Our gratitude is also directed toward 
Sharon Morsink for a guided tour of {\tt rns}, and Nikolaos Stergioulas for 
making his code available. Finally, we thank an anonymous referee for pointing 
out key references and suggesting ways to improve the manuscript.

\end{document}